\documentclass[12pt]{article}
\usepackage{times}
\usepackage{geometry}
\usepackage[utf8]{inputenc}
\usepackage{enumitem,amssymb}
\usepackage{ragged2e}
\usepackage{natbib}
\usepackage{graphics,graphicx}
\usepackage{wrapfig}
\setcitestyle{aysep={}} 
\newlist{thematic}{itemize}{8}
\setlist[thematic]{label=$\square$}
\usepackage{pifont}

\usepackage[colorlinks,linkcolor=blue,citecolor=black]{hyperref}
\hypersetup{urlcolor=blue}

\begin{document}
\raggedright
\huge
Astro2020 Science White Paper \linebreak

High Angular Resolution Astrophysics:  Resolving Stellar Surface Features \linebreak
\normalsize

\noindent \textbf{Thematic Areas:} \hspace*{60pt} $\square$ Planetary Systems \hspace*{10pt} $\square$ Star and Planet Formation \hspace*{20pt}\linebreak
$\square$ Formation and Evolution of Compact Objects \hspace*{31pt} $\square$ Cosmology and Fundamental Physics \linebreak
  $\boxtimes$  Stars and Stellar Evolution \hspace*{1pt} $\square$ Resolved Stellar Populations and their Environments \hspace*{40pt} \linebreak
  $\square$    Galaxy Evolution   \hspace*{45pt} $\square$             Multi-Messenger Astronomy and Astrophysics \hspace*{65pt} \linebreak
  
\textbf{Principal Author:}

Name:	Rachael M.\ Roettenbacher
 \linebreak				
Institution:  Yale University
 \linebreak
Email: rachael.roettenbacher@yale.edu
 \linebreak
Phone:  203-436-9621
 \linebreak
 
\textbf{Co-authors:} 

Ryan P.\ Norris (Georgia State University)

Fabien Baron (Georgia State University)

Kenneth G.\ Carpenter (NASA GSFC)

Michelle J.\ Creech-Eakman (New Mexico Tech)

Douglas Gies (Georgia State University)

Thomas Maccarone (Texas Tech University)

John D.\ Monnier (University of Michigan)

Gioia Rau (NASA GSFC)

Stephen Ridgway (NOAO)

Gail H.\ Schaefer (The CHARA Array/Georgia State University)

Theo ten Brummelaar (The CHARA Array/Georgia State University)
\linebreak

\textbf{Abstract  (optional):}

\justifying 

\noindent{}We are now in an era where we can image details on the surfaces of stars.  When resolving stellar surfaces, we see that every surface is uniquely complicated.  
Each imaged star provides insight into not only the stellar surface structures, but also the stellar interiors suggesting constraints on  evolution and dynamo models.  As more resources become operational in the coming years, imaging stellar surfaces should become commonplace for revealing the true nature of stars.  Here, we discuss the main types of stars for which imaging surface features is currently useful and what improved observing techniques would provide for imaging stellar surface features.

\pagebreak

\section{Introduction}

\begin{wrapfigure}[18]{4}{7.75cm}
\begin{center}
\vspace{-1.cm}
%\hspace{-1.8cm}
\includegraphics[scale=0.27]{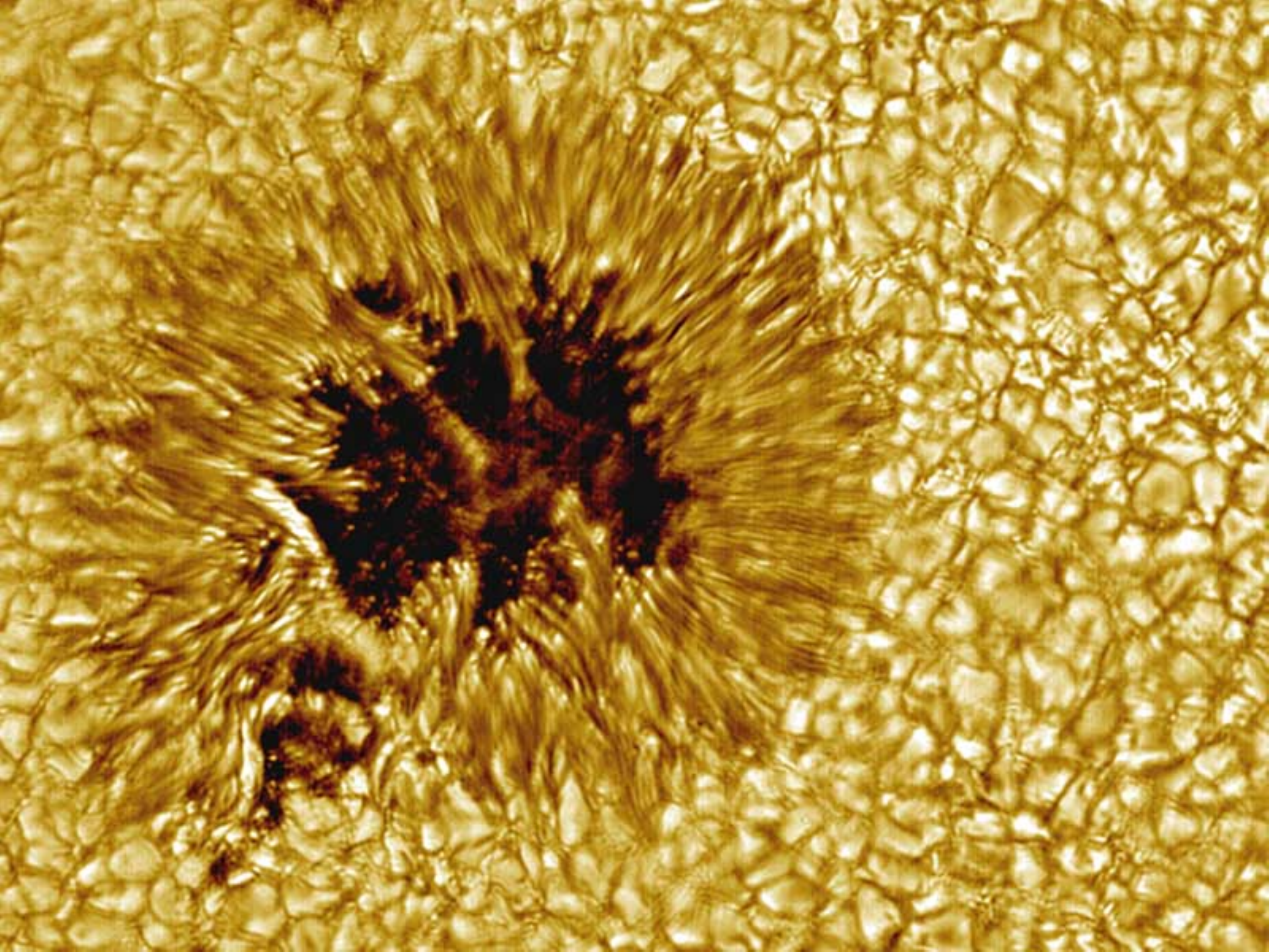}
\vspace{-.4cm}
\caption{Detail of a small portion of the solar surface showing solar convective cells and a sunspot.  Analogous features to these can be resolved on other stars with long-baseline optical interferometry.  Credit:  APOD/Vacuum Tower Telescope/NSO/NOAO  }
\vspace{-0.25cm}
 \line(1,0){220}
\label{fig:sun}
\end{center}
\end{wrapfigure}From direct observations, we have long known that the solar photosphere is not featureless, but covered with evolving sunspots and convective cells \citep[written documentation of sunspots dates back to 200 BC; see review by][see also, Figure \ref{fig:sun}]{vaq09}.  In the outer solar envelope, convection is the most efficient form of energy transport.  In each convective cell, hot material rises toward larger radii and  the material that has cooled sinks back towards the center of the Sun.  Solar, convective cells in the photosphere are expected to have lifetimes on the order of minutes \citep[e.g.,][]{ali87}.  When the efficient convective transport is stifled by strong, localized magnetic fields, the surface is cooler by up to  $\approx 1000$ K.  This cool region, or sunspot, will look dark against the stellar photosphere. 

On stars beyond the Sun, variations due to surface structures rotating in and out of view were first identified as similar to sunspots when \citet{kro47} observed variability in the light curve of AR Lacartae.  Theories of the convection of supergiants predict large temperature discrepancies across the stars' convective cells; these features  are not easily detected due to the stars' slow rotation.  
Even though stars are too distant for the straight-forward photographing we can do for the Sun, a number of methods can be used to reconstruct the stellar surface.  However, the only method capable of directly imaging the star as it appears on the sky is interferometry.  

In order to resolve a star in visible or infrared wavelengths, today's single-aperture telescopes are not sufficient for measuring most stellar diameters and a multiple-telescope, long-baseline optical interferometer is required.  Interferometers are now capable of measuring more than just stellar diameters, revealing features on a stellar surface, which requires extensive observation (the distance difference stellar light travels between each pair of telescopes---the projected baseline---maps to a specific point on the star; $uv$ coverage).  To account for the gaps in observational coverage, aperture synthesis imaging is used.  This technique transforms the brightness variations and (closure-phase) asymmetries into an image of the stellar surface.  With the ability to image the surface, we are  able to \emph{directly} measure stellar inclination, rotation, distortion from a sphere, and detect brightness asymmetries of the photosphere.

Here, we discuss the stellar surface features that we can currently directly image--convective cells and starspots.  Additionally, we discuss future prospects for these studies with more advanced interferometric capabilities.

\section{Convection and Granulation}
For both low- and high-mass stars, the late stages of stellar evolution entail a large increase in radius, significant mass-loss, and notable variability. Understanding the link between mass-loss, variability, and surface activity in these stars has long been a goal of stellar astrophysics to better describe the late stages of stellar evolution.
In the 1970s, Martin Schwarzschild used mixing length theory to suggest that large, evolved stars exhibit convection on a scale significantly larger than that of the Sun \citep{sch1975}. Indeed, simulations of evolved stars using 3D radiative hydrodynamics show surfaces with large ($\sim 1 R_{*}$), long-lived convection cells, smaller, shorter-lived  granulation-like features, and a surface whose appearance is highly wavelength dependent \citep{chia2009,chia2010}. These features can bias measurements of stellar parallax, radial velocity, chemical abundance, and fundamental stellar parameters \citep{chia2011}. Although photometry \citep{kiss2006} and spectroscopy  \citep{gray2008,krav2018} have offered tests of these 3D models, interferometric imaging of stellar surfaces is one of the most powerful ways to study the scale, lifetime, and variety of surface features in evolved stars.  Evolved red supergiants (RSGs) can serve as probes of the chemical evolution of other galaxies \citep{gazak2015}, and their high luminosity in the near-infrared makes them ideal targets for studies of distant galaxies with the James Webb Space Telescope  \citep[JWST;][]{lev2018}.  Understanding the impact of surface features on stellar parameter determinations is going to become ever more important in the coming decade.  

Simulations suggest that the surface of RSGs exhibit a few very large convection cells with lifetimes on the order of 1000s of days, within which are smaller granulation-like features with lifetimes of 100s of days \citep{chia2014}. Spectropolarimetry suggests that these cells are associated with the weak magnetic field found in these stars and that plasma in these cells exhibit horizontal and vertical velocities on the order of 20 km/s \citep{lp2018}. Although evidence suggests that convection alone cannot explain the origin of mass-loss in RSGs, it is unclear whether convection is a contributing factor to mass-loss. Observations with long-baseline interferometers such as the Very Large Telescope Interferometer (VLTI) and Center for High Angular Resolution Astronomy (CHARA) Array, show surfaces with several bright features at the same scale as the predicted granulation \citep{baron2014,mont2018}. Long-term, time-series studies of RSGs are underway, in hopes of comparing the timescale of convective turn-over and measuring the properties of surface features for comparison to models. 

As a result of their lower gravity, convection in asymptotic giant branch (AGB) stars is even stronger than in RSGs, reaching into the core. Simulations of AGBs with 3D radiation hydrodynamics (RHD) models also show a highly irregular surface, with large features of sharp contrast \citep{freytag2017}. The most detailed images of an AGB star to date were produced from observations of  $\pi^{1}$ Gruis using the VLTI by \citet{paladini2018}. These images show a surface with several large features and a strong wavelength dependence. As with RSGs, long-term, time-series studies of AGBs are underway in order to better characterize the behavior of these features and their impact on stellar parameter determinations, as well as convection's contribution to stellar behavior such as mass-loss.

\newpage 
\section{Magnetic and Abundance Spots}

Stars showing  modulation from ``starspots'' rotating in and out of view include RS Canum Venaticorum (RS CVn), BY Draconis (BY Dra), and peculiar A (Ap) stars.  These stars vary dramatically in their evolutionary states and the mechanisms behind their spots.  

RS CVn stars are typically binary systems where the primary star is  evolved  (e.g., giant or subgiant) and the companion is less-evolved (e.g., subgiant or main-sequence). These stars are typically tidally-locked  with orbits often between 1-3 weeks and with the primary star showing strong indications of stellar activity through rotational modulation and/or activity indicators like H$\alpha$ and Ca II H\&K emission.  
These starspots are believed to be analogous to sunspots and are formed by dense magnetic fields jutting out of the surface stifling convection and preventing efficient energy transport.  The starspots of RS CVn stars tend to be much larger than sunspots, likely due to the larger convective zones on the giant stars and rotational spin-up from the close companion.  Similar to RS CVns, BY Dra stars are typically single, main-sequence stars that show similar signs of activity, and are expected to form spots in the same way.

Ap stars show peculiar chemical abundances where an increased metal content is concentrated at the poles of the stellar magnetic field.  The enhanced metallicity in these regions works to absorb photons making the area appear darker.  Ap stars can have offset rotational and magnetic poles allowing for rotational modulation and the photometric detection of these spots.  These stars have strong magnetic fields making them particularly interesting for Zeeman Doppler imaging (ZDI) studies.   ZDI reconstructs stellar surfaces from the perturbations in the Stokes polarization profiles and are indicative of the structure of the stellar magnetic field \citep[e.g.,][]{ros15}.

\begin{wrapfigure}[34]{r}{16.5cm}
\begin{center}
\vspace{-1.25cm}
\hspace{-1.cm}
\includegraphics[scale=0.55]{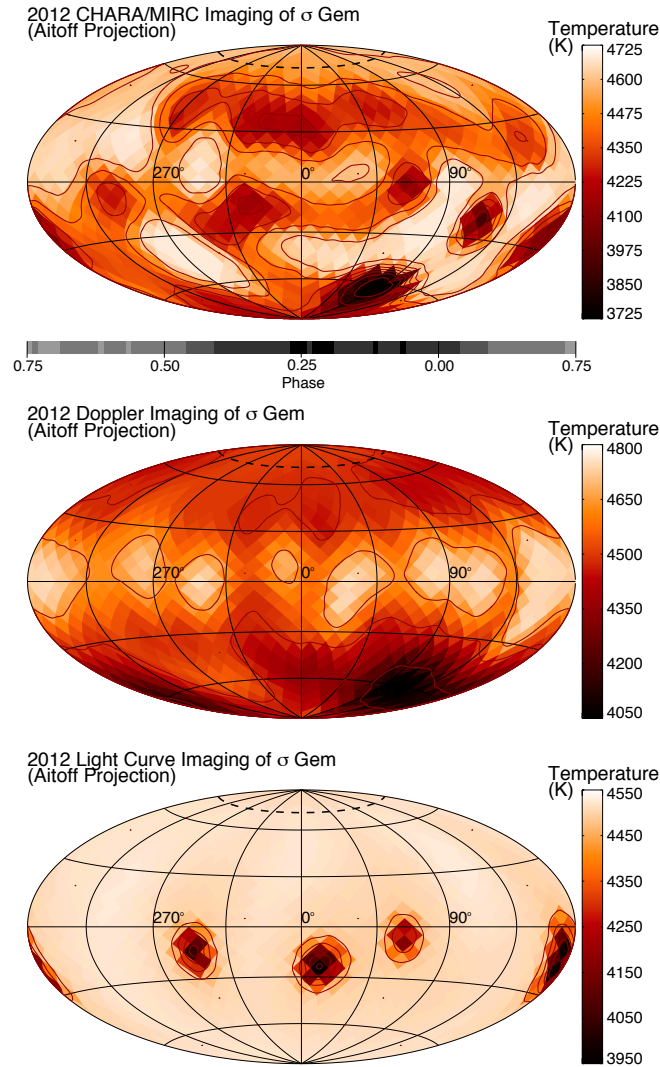}
\vspace{-1.0cm}
\caption{Aitoff projections of temperature maps for three imaging techniques of 2012 data of the spotted giant star $\sigma$ Geminorum.  The region above the dashed line on the top pole is unobserved due to inclination.  Top:  Interferometric aperture synthesis imaging with a color bar below indicating the number of times each phase was observed (lightest gray indicates three nights and darkest gray indicates seven nights).  Middle:  Doppler imaging map.  Bottom:  Light-curve inversion imaging map. Reprinted from \citet{roe17}.  
\emph{For an animated image of how $\sigma$ Gem appeared on the sky in H-band during late 2012, see \url{https://rmroettenbacher.github.io/Roettenbacheretal2017.gif}. } }
\vspace{-0.25cm}
 \line(1,0){470}
\label{fig:siggem}
\end{center}
\end{wrapfigure}

RS CVns, with their comparatively large starpots are the stars observed most frequently for imaging studies.  In addition to interferometric imaging, RS CVn starspots have been indirectly imaged with light-curve inversion and Doppler imaging methods \citep[e.g.,][and many more]{roe11, jar18}.  Because each imaging technique has a particular set of criteria that must be met in order to obtain results,  only a few stars have been imaged with more than one technique.  Here, to illustrate the imaging of starspots, we include the example of $\sigma$ Geminorum, an RS CVn system that has been imaged with the three different imaging techniques (see Figure \ref{fig:siggem}).

In addition to individual snapshots of stellar surfaces, sequential observations can be used to understand spot evolution, spot characteristics, and surface behaviors, such as differential rotation \citep[with non-interferometric observations, e.g.,][]{hat98,col02,roe13}.  In particular, some studies have shown evidence of differential rotation in the opposite sense of that seen on the Sun, that is with the equatorial material rotating more slowly than the polar material \citep{kov15}.  Understanding the evolution of starspots over time by repeated observations and differential rotation measurements as observed with photometry and spectroscopy is limited due to latitude degeneracies, in particular.  However, a series of interferometric images would provide new insights into surface and dynamo evolution.  

In longer (radio) wavelengths, the extended photosphere of giant stars, like  RS CVns, can be observed to reveal structures and features.  With the future increased sensitivity of the Next Generation Very Large Array (ngVLA; at 85 GHz, sensitivity of 1000 K in one hour at 0.4 mas angular resolution),  stellar chromospheres will be resolved at radio wavelengths for direct comparison with photospheric active regions mapped through optical interferometry \citep[as in, e.g.,][]{cari18}.  The structure of stellar winds will also be resolved \citep[as in, e.g.,][]{mac18}.    Ultraviolet  

\ 

\ 

\ 
 
 \ 
 
 \ 
 
 \ 
 
 \ 
 
 \ 
 
 \ 
 
 \ 
 
 \ 
 
 \ 
 
 \ 
 
 \ 
 
 \ 
 
 \ 
 
 \

 \  
 
\ 

\ 

 \ 
 
 \ 
 
 \ 
 
 \ 
 
 \ 
 
 \ 
 
 \ 
 
  \ 
  
  \ 
  
  \

 \ 
 
 \ 
 
 \

 \ 
 
  \ 
  
\ 
  
  \noindent{}wavelengths can also reveal chromospheric information on the shape of the magnetic field \citep[e.g.,][]{car18,rau18}.

\section{Advancements}

Stellar astrophysics has had a recent rejuvenation with advancements in high-cadence, space-based photometry, extreme-precision radial velocities, and long-baseline optical interferometry.  These resources can be used in concert to provide better understandings of stellar astrophysics.  
Continuing to move forward, these and new resources will be utilized to further understand the structure and evolution of stars and their magnetic fields.

With the recent launch of the Transiting Exoplanet Survey Satellite (\emph{TESS}) and the upcoming CHaracterising ExOPlanet Satellite (CHEOPS) and PLAnetary Transits and Oscillabions of stars (PLATO) missions, space-based photometry will continue to flourish while searching for transiting planets.  These satellites will observe stars for extended periods of time obtaining high-cadence observations that will allow for detailed observations of stars changing on timescales ranging from minutes to weeks.  

The next generation of extreme precision radial velocity spectroscopic instruments are coming online now (e.g., EXtreme PREcision Spectrometer (EXPRES), Echelle SPectrograph for Rocky Exoplanets and Stable Spectroscopic Observations
(ESPRESSO), Potsdam Echelle Polarimetric and Spectroscopic Instrument (PEPSI)), and will provide not only the radial velocity precision to detect an Earthlike planet around a Sunlike star, but will provide high signal-to-noise, high-resolution observations that will routinely allow for detailed mappings of starspot features and careful measurements of photospheric temperature fluctuations.  These observational resources will provide crucial comparative and supplementary observations and tests to detailed interferometric imaging campaigns.  

\begin{wrapfigure}[15]{4}{7.75cm}
\begin{center}
\vspace{-1.5cm}
%\hspace{-1.8cm}
\includegraphics[scale=0.27]{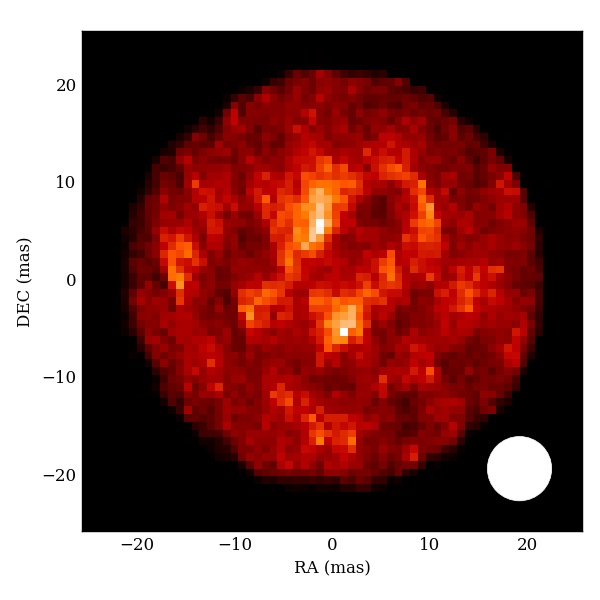}
\vspace{-.4cm}
\caption{Simulation of a RSG imaged using MROI, based on  a 3D RHD simulation from \citet{chia2010} and one of the planned configurations for the interferometer.   }\vspace{-0.25cm}
 \line(1,0){220}
\label{fig:rsg}
\end{center}
\end{wrapfigure}Current interferometers (the CHARA Array and VLTI) and those being upgraded or under construction (Navy Precision Optical Interferometer (NPOI) and Magdalena Ridge Observatory Interferometer (MROI), respectively; see Figure \ref{fig:rsg} for a simulated observation of Betelgeuse using a planned MROI telescope configuration) will continue to provide an unmatched view of stellar surfaces with submilliarcsecond resolution.  While the number of stars that can be imaged with such interferometers is limited, those stars that can be studied provide critical tests to stellar evolution and magnetic dynamo theories.

Further advances in interferometry will offer the most dramatic improvements in understanding stellar activity and evolution.  Future interferometers with even longer baselines and more numerous telescopes will provide the means to observe features on stellar surfaces at higher resolution, allowing for improved detection of convective cells and starspots, while allowing for the potential access to detection of brightness variations due to other phenomena, such as non-radial pulsations and direct observations of planets in transit.  With more telescopes and a variety of available baselines, it will become possible to study stellar surfaces at scales both large and small, thus providing a stringent test of theoretical predictions and a better understanding of how surface features impact measurements of other stellar properties---a vital step as missions like JWST begin to study evolved stars beyond the galaxy. Moreover, any advances to stellar imaging capabilities will lead directly to updating stellar evolution models \citep[e.g.,][]{mar13,mar17}, which will also affect not only stellar astrophysics, but will improve stellar population models and our greater understanding of the universe.

\pagebreak

\end{document}